\documentclass[9pt,twocolumn,twoside]{opticajnl}

\usepackage{siunitx}
\usepackage{textcomp}
\usepackage{appendix}
\usepackage{chemformula}
\usepackage{cleveref}
\Crefname{equation}{Eq.}{Eqs.}
\crefname{pluralequation}{Eqs.}{Eqs.}
\Crefname{figure}{Fig.}{Figs.}
\crefname{pluralfigure}{Figs.}{Figs.}
\Crefname{tabular}{Tab.}{Tabs.}

\DeclareMathOperator{\sinc}{sinc}

\journal{opticajournal}

\setboolean{shortarticle}{true}

\title{Third-harmonic generation in silica wedge resonators}

\author[$\dagger$,1, 2]{Jorge H. Soares}
\author[$\dagger$,1, 2]{La\'is Fujii dos Santos}
\author[1,2]{Felipe G. S. Santos}
\author[1,2]{Marvyn Inga}
\author[1,2]{Yovanny A. V. Espinel}
\author[1,3]{Gustavo S. Wiederhecker}
\author[1,2,*]{Thiago P. Mayer Alegre}

\affil[1]{Photonics Research Center, University of Campinas, Campinas, SP, Brazil}
\affil[2]{Applied Physics Department, Gleb Wataghin Physics Institute, University of Campinas, Campinas, SP, Brazil}
\affil[3]{Quantum Electronics Department, Gleb Wataghin Physics Institute, University of Campinas, Campinas, SP, Brazil}
\affil[$\dagger$]{These authors contributed equally}
\affil[*]{alegre@unicamp.br} 

\begin{abstract}
Whispering-gallery-mode microcavities are known to have high optical quality factor,  making them suitable for nonlinear optical interactions. Here, third-harmonic generation is observed using a relatively small radius wedge silicon oxide optical microcavity. The small radii wedge microdisks can be dispersion-tailored to obtain either normal or anomalous group velocity dispersion. In our case, we operate in the normal dispersion regime preventing frequency comb generation by suppressing infrared cascading four-wave mixing. This allows for a clean third harmonic generation at phase-matched visible optical modes. Tunability of the third-harmonic emission is obtained due to a combination of the thermo-optical and Kerr effects. An additional thermal control of the phase matching condition allows for optimizing third-harmonic generation, and agreement between this process and couple-mode theory is demonstrated.
\end{abstract}

\setboolean{displaycopyright}{true}

\begin{document}

\maketitle

\section{Introduction}
\label{sec:intro}

\begin{figure}[ht!]
	\centerline{\includegraphics[scale=1]{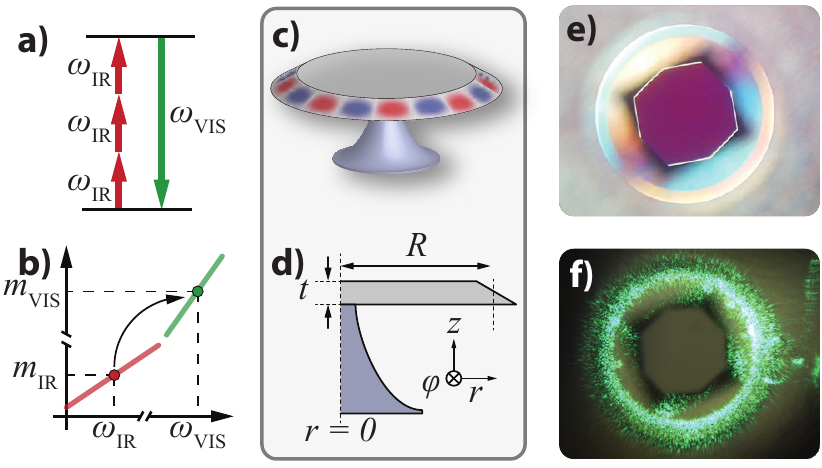}}
	\caption{ 
    \textbf{a)} Energy level diagram of the third harmonic process. VIS and IR: visible and infrared frequencies.
	\textbf{b)} Phase-matching condition for THG: $m_\text{VIS}(\omega_\text{VIS})=3 m_\text{IR}(\omega_\text{IR})$.
    \textbf{c)} Illustration of the wedge microdisk, with an illustration of the optical mode apparent at the wedged surface.
	\textbf{d)} Transverse profile of the wedge disk, with dimensions and coordinates.
	\textbf{e)} Optical microscopy image of the fabricated sample (top view).
	\textbf{f)} Microscope image of visible (green) light generated through THG.}
	\label{fig:cav}
\end{figure}

Optical microcavities are well known to provide high optical confinement and quality factors~\cite{Vahala2003OpticalMicrocavities, Armani2003}. New micro-fabrication techniques have driven such systems to a new frontier, where photon lifetime exceeds a microsecond~\cite{Lee2012ChemicallyChip, Ji2017, Luke2013}. This allows for a large intracavity photon number enabling non-linear optics at moderate pump powers. In particular, ultra-high optical quality factors (exceeding $800$ million)~\cite{Lee2012ChemicallyChip} were reported in large radii wedge microdisks based upon a silicon-oxide substrate (depicted in \Cref{fig:cav}(\textbf{a,b}). Those devices exhibited anomalous group velocity dispersion (GVD) \cite{Lee2017} and were used to demonstrate frequency comb generation \cite{Li2012, Lee2017}, stimulated Brillouin lasers \cite{Lee2012ChemicallyChip} and erbium-doped on-chip lasers \cite{Kippenberg2006}. We have recently demonstrated~\cite{Fujii2020} that small radii wedge microdisks can also be dispersion-tailored to obtain either normal or anomalous  GVD. At small radii, angular momentum contributes to a larger degree of normal GVD, but the control of the wedge angle can compensate for it.

Here we demonstrate tunable third-harmonic-generation (THG) in such small radius wedge microdisks. Although THG was reported in multiple silica-based  resonators~\cite{Farnesi2014, Carmon2007, Lee2013}, the ability to control the GVD and operate in the normal dispersion regime prevents frequency comb generation, since cascading four wave mixing of the infrared pump is suppressed. Therefore, clean third harmonic generation at phase-matched visible optical modes is achieved. Furthermore, we explore a thermal tuning technique~\cite{guoSecondharmonicGenerationAluminum2016, luPeriodicallyPoledThinfilm2019, Surya2018EfficientMicrorings, bruch17000Secondharmonic2018, huAllopticalThermalControl2020} to control the phase-matching condition in order to enhance the frequency conversion efficiency.

\begin{figure*}[ht!]
	\centerline{\includegraphics[scale=1]{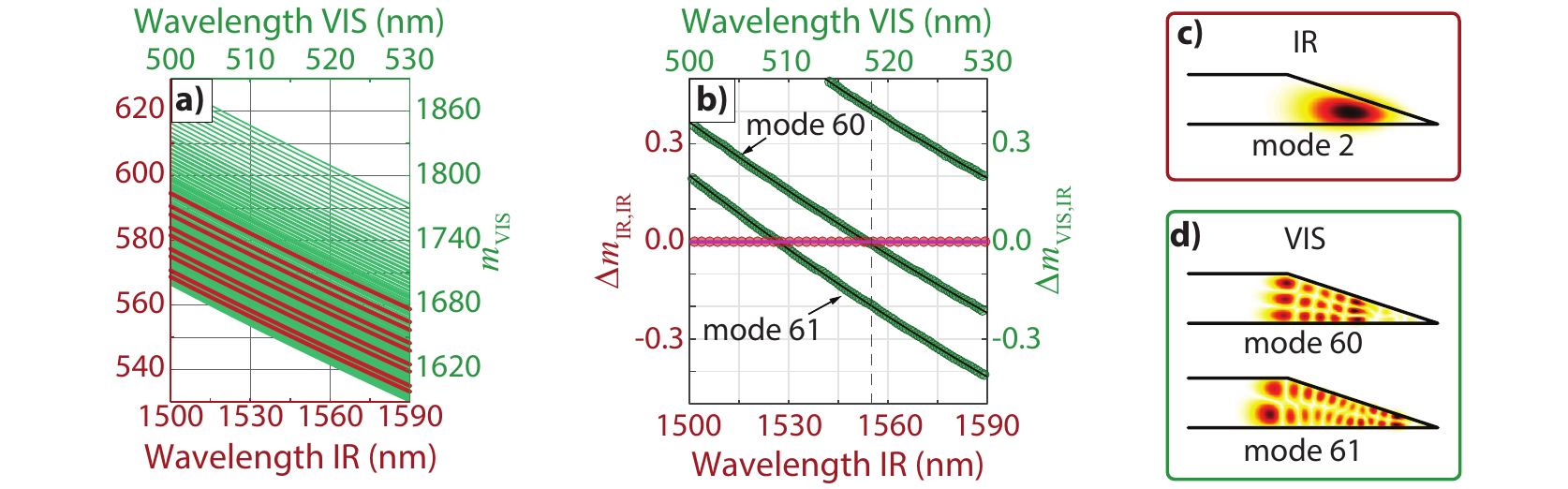}}
	\caption{
	\textbf{a)} Comparison between the simulated VIS and IR mode numbers at the corresponding spectral windows.
	\textbf{b)} Residual dispersion curves highlighting the fundamental TM IR mode. The $m_\text{IR}(\lambda_\text{IR})$ values of this mode were subtracted from all the IR curves, whereas $3m_\text{IR}(\lambda_\text{VIS})$ were subtracted from the VIS curves.
	\textbf{c)} Transverse field profile of the fundamental TM mode (labeled 2) at 1550 nm.
	\textbf{d)} Transverse field profile of higher order modes (labeled 60 and 61) at 517 nm.}
	\label{fig:pm}
\end{figure*}

Third-harmonic-generation (THG) is a frequency-conversion process in which the converted optical frequency is three times that of the input laser beam~\cite{boydNonlinearOptics2008}. This process is mediated by the third order nonlinear susceptibility of the medium ($\chi^{3}$) and therefore involves four photons: three infrared (IR) pump photons and a single visible (VIS) photon, as illustrated in the energy diagram of \Cref{fig:cav}\textbf{a)}.

Besides energy conservation, the process is also bounded to momentum conservation; in other words, ensuring phase-matching is crucial to observe efficient conversion. In axisymmetric optical cavities, the periodic boundary conditions at the propagation direction ($\varphi$ direction in \Cref{fig:cav}\textbf{c)}) forces the substitution of the wave number by the azimuthal mode number $m$. Thus, the phase-matching condition reads $m_\text{VIS}=3 m_\text{IR}$.

Due to GVD, it is not possible to satisfy the phase-matching condition with modes of the same transverse family, as exemplified by the slope of the $m$ vs. $\omega$ curve at VIS and IR frequencies illustrated in \Cref{fig:cav}\textbf{b)}. The advantage of whispering gallery resonators is that they support multiple higher order modes; these higher order modes have a lower wavenumber due to their transverse profile, increasing the chances of phase-matching between a low transverse order IR mode with a higher transverse order VIS mode.

\Cref{fig:pm}\textbf{a)} shows the result of a Finite Element Method (FEM) simulation of the dispersion for the first few IR (red) and VIS (green) modes for a wedge cavity of radius \SI{100}{\um}, thickness \SI{3}{\um} and side angle \SI{18.5}{\deg}. The large density of high order VIS modes is seen as a dark green region in this picture. It is also clear that phase-matching is only possible with higher order visible modes. To identify the visible mode order satisfying phase-matching, \Cref{fig:pm}\textbf{b)} shows the simulated difference in VIS and IR azimuthal mode numbers as a function of wavelength. The chosen reference mode ($\Delta m_\text{IR,IR} = m_\mathrm{IR}(\lambda_1) - m_\mathrm{IR}(\lambda_1) = 0$ where $\lambda_1$ is the mode's wavelength) is the fundamental TM IR mode illustrated in \Cref{fig:pm}\textbf{c)}, whereas $\Delta m_\text{VIS, IR} = m_\mathrm{VIS}(\lambda_1/3) - 3 m_\text{IR}(\lambda_1)$, where $m_\mathrm{VIS}(\lambda_1/3)$ is a linear interpolation of $m_\mathrm{VIS}$ at the third harmonic wavelength $\lambda_1/3$, is shown for a few VIS modes. Phase-matching is satisfied whenever a green dot coincides with the red one (in our case, for a pump at \SI{1550}{\nm}, the reference IR mode phase-matches mode 60 in the VIS range (\Cref{fig:pm}\textbf{d)}) as indicated by the black dashed line in \Cref{fig:pm}\textbf{b)}. On the other hand, the difference in transverse IR and VIS profiles reduces their spatial overlap and the efficiency of the THG process~\cite{Rodriguez2007}.

As a final introductory remark, it is worth taking a closer look at the dynamics of this system. The coupled-mode equations for doubly-resonant THG are~\cite{Rodriguez2007},
\begin{equation} \label[pluralequation]{eq:cmt}
\begin{aligned}
    \dot{a}_1 = & -(i (\Delta_1 + \alpha |a_1|^2 + 2\alpha |a_3|^2) + \gamma_1) a_1 - i g (a_1^*)^2 a_3 + \sqrt{2 \theta_1} s_\mathrm{in}, \\
    \dot{a}_3 = & -(i (\Delta_3 + 2\alpha |a_1|^2 + \alpha |a_3|^2) + \gamma_3) a_3 - i g a_1^3,
\end{aligned}
\end{equation}
where $a_1$ ($a_3$) is the IR (VIS) amplitude corresponding to a mode of resonant frequency $\omega_1$ ($\omega_3$), total loss rate $\gamma_1$ ($\gamma_3$) and detuning $\Delta_1 = \omega_1 - \omega_L$ ($\Delta_3 = \omega_3 - 3\omega_L$), $\omega_L$ being the frequency of the pumped wave. The strength of the optical Kerr effect is given by the coefficient $\alpha$, being multiplied by an extra factor of two in the case of cross-phase modulation~\cite{boydNonlinearOptics2008}; $g$ describes the strength of the nonlinear interaction responsible for THG, $\theta_1$ is the coupling rate between the cavity and the IR waveguide and $s_\mathrm{in}$ is the input wave such that its squared absolute value is the pump power $P_\mathrm{in}$. Under the undepleted pump approximation, the stored energy at steady-state reads
\begin{equation} \label[pluralequation]{eq:energy}
\begin{aligned}
    |a_1|^2 = & \frac{ 2 \theta_1 P_\mathrm{in} }{ (\Delta_1 + \alpha |a_1|^2)^2 + \gamma_1^2 }
    \\
    |a_3|^2 = & \frac{ g^2 |a_1|^6 }{ (\Delta_3 + 2\alpha |a_1|^2)^2 + \gamma_3^2 }
\end{aligned}
\end{equation}

Besides showing that the generated VIS power is proportional to $P_\mathrm{in}^3$, \Cref{eq:energy} display another interesting feature: because the (power-dependent) nonlinear frequency shift of the IR mode (caused by self-phase modulation) is half that of the VIS mode (due to cross-phase modulation), the optical Kerr effect may be able to compensate deviations from the condition $\omega_3 = 3\omega_1$ and therefore enhance THG. As we show experimentally, this ``dynamical phase-matching compensation'' is crucial to compensate GVD and therefore lead to \emph{tunable} THG.

\section{Experiment}

Our optical cavities were fabricated through a two step wet etching process as described in Ref.~\cite{Fujii2020}, a process inspired by \cite{Lee2012ChemicallyChip}, but using a wet etch for both the silica device layer (thickness $t=\SI{3}{\um}$) and the underlying silicon pedestal, the resulting device geometry is illustrated in \Cref{fig:cav}.

\begin{figure}[htb]
	\centerline{\includegraphics[scale=1]{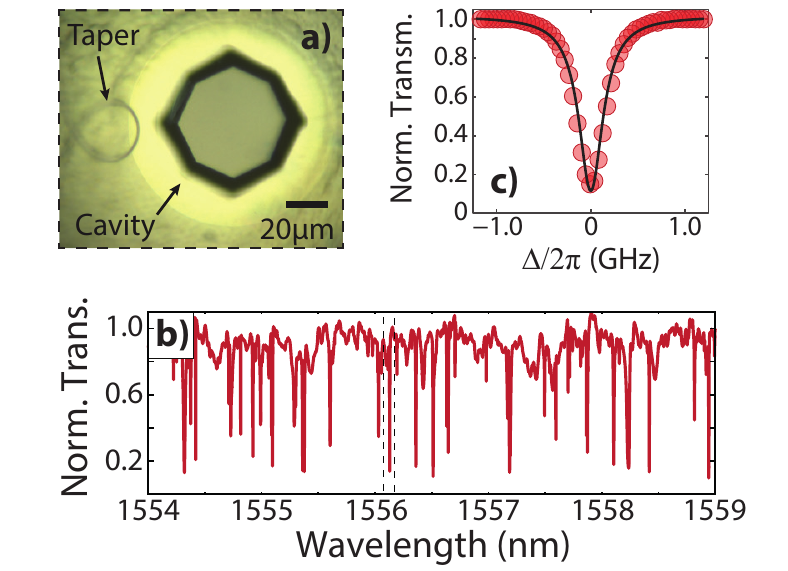}}
	\caption{\textbf{a)} Microscope picture of the taper coupled to the cavity.
	\textbf{b)} Cavity transmission measured as a function of IR pump wavelength. The highlighted resonance (centered at \SI{1553.13}{\nm}) is plotted in \textbf{c)} (red dots), with the Lorentzian fit in black.}
	\label{fig:setup}
\end{figure}

Light is coupled into and out of the device via a tapered optical fiber~\cite{Ding:2010vs}, which is placed at the near-field of the microdisk, as shown in \Cref{fig:setup}\textbf{a)}. For wavelengths sufficiently close to cavity resonances, the taper's guided mode overlaps with the resonator's evanescent field and leads to energy exchange. A linear characterization of the microdisk is performed by measuring the transmission spectrum of the cavity, pumped by a tunable IR laser, resulting in the broadband spectrum shown in \Cref{fig:setup}\textbf{b)}. Fitting a Lorentzian curve to the resonance at \SI{1553.13}{\nm} (\Cref{fig:setup}\textbf{c)}), we obtain a loaded optical quality factor of $Q=6.2\times 10^5$. A more detailed study of the $Q$ values throughout the IR spectrum is provided elsewhere~\cite{Fujii2020}. Notice that mode identification from the transmission spectrum is hindered by the large number of excited modes; however, from the dispersion study conducted in Ref.~\cite{Fujii2020}, we know the average free spectral range to be approximately \SI{3}{\nano\meter}.

\begin{figure}[htb]
	\centerline{\includegraphics[scale=1]{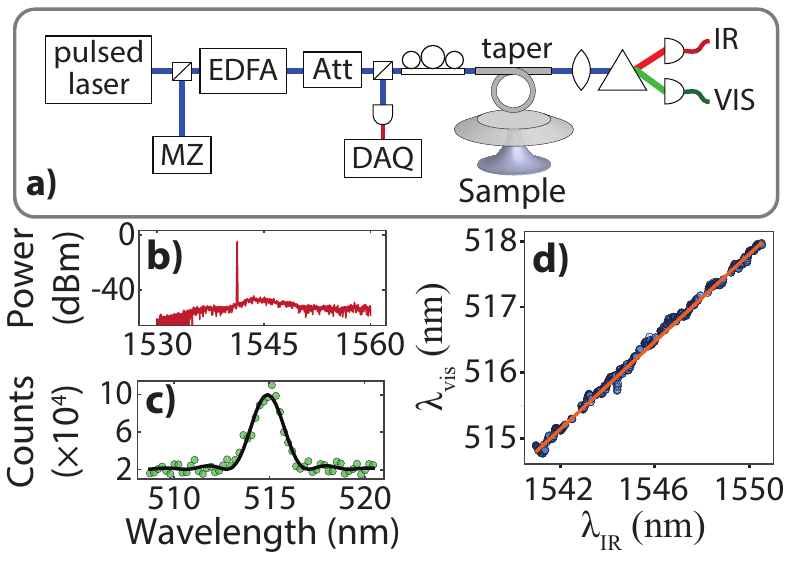}}
	\caption{\textbf{a)} Experimental setup: Laser light source; MZ: Mach-Zehnder interferometer for wavelength calibration; Erbium doped fiber amplifier (EDFA); VIS and IR: visible and infrared optical detectors; DAQ: digital analog converter. Att: optical attenuator.
	\textbf{b)} IR transmission spectrum captured with an optical spectrum analyser (OSA).
	\textbf{c)} VIS emission captured with a spectrometer.
    \textbf{d)} Wavelength corresponding to third harmonic peak, $\lambda_\text{VIS}$, versus IR resonance $\lambda_\text{ IR}$. Slope of linear fit is 3.}
	\label{fig:setup_thg-lambda}
\end{figure}

To perform the nonlinear experiment, we modulated the pump laser and then amplified the optical pulses with an Erbium-doped fiber amplifier (EDFA), increasing the input peak power to a few Watts while ensuring a quasi-CW regime by maintaining the pulse duration (\SI{400}{\ns}) longer than the lifetime of the photons in the cavity (less than \SI{1}{\ns}). The maximum power achieved at the EDFA's output (\SI{7.1}{\W} peak power, \SI{460}{\mW} average power) can be varied by controlling the attenuation prior to the taper's input as illustrated in \Cref{fig:setup_thg-lambda}\textbf{c)}. The IR and VIS output fields are separated by a prism and measured with various instruments according to the experimental demands. Despite the distinct bright green scattered by the cavity (\Cref{fig:cav}\textbf{d)}), we demonstrate THG using this setup to observe the spectral signature of the collected visible light as well as how its power depends on the input IR power.

\begin{figure*}[htb!]
	\centerline{\includegraphics[scale=1]{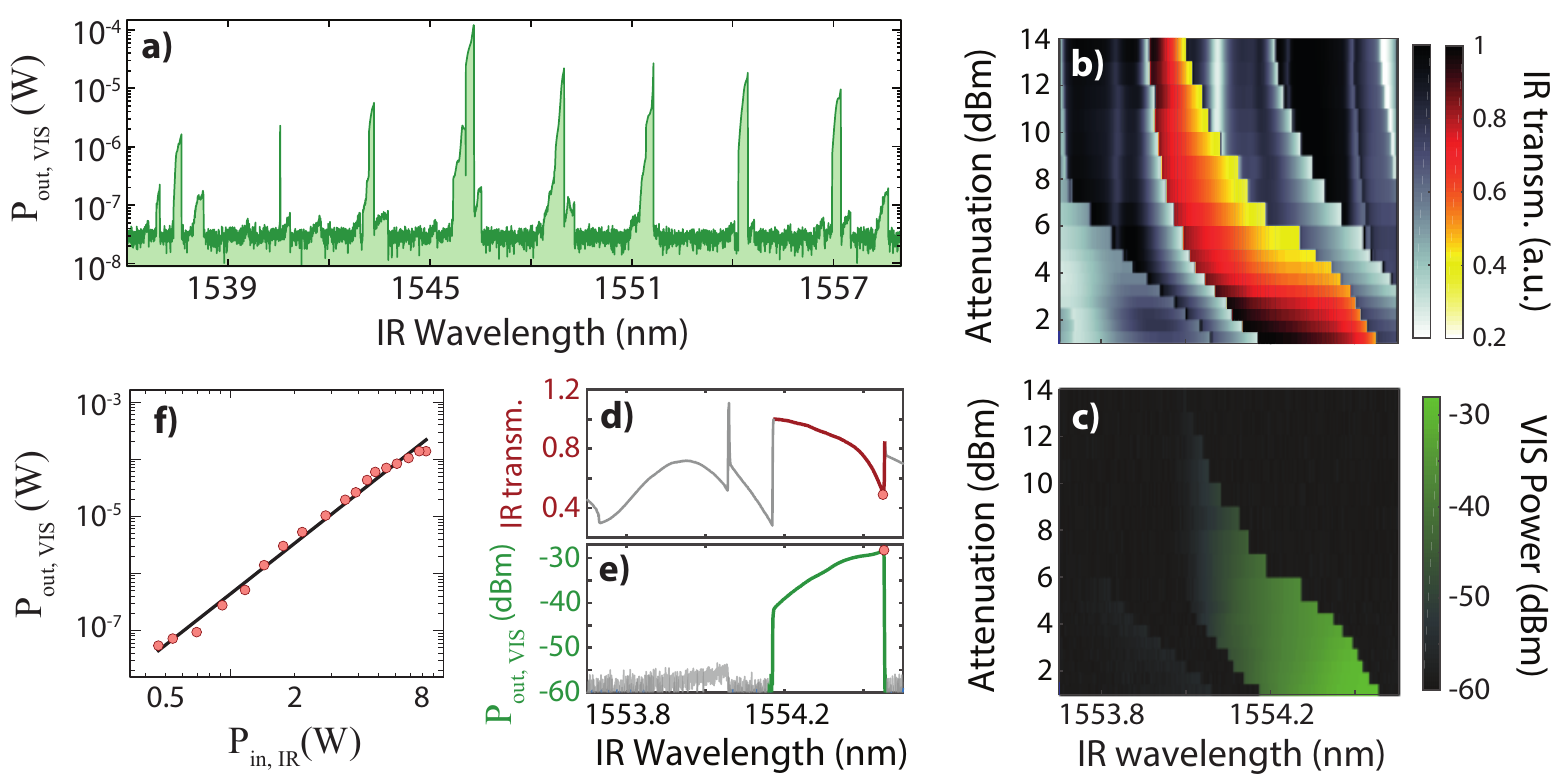}}
	\caption{ \textbf{a)} Power of the VIS transmission as a function of IR pump wavelength.
    \textbf{b)} and \textbf{c)}: IR and VIS transmission spectra for varying input attenuation and IR pump wavelength. The mode highlighted in red in b) is responsible for the TH signal in c).
    \textbf{d)} and \textbf{e)}: IR and VIS transmission spectra for zero attenuation (highest input power).
	\textbf{f)} Measured visible power $P_\text{out, VIS}$ as a function of input IR power $P_\text{in, IR}$. These values were taken at wavelengths that maximize the TH emission, as exemplified in d) and e).
}
	\label{fig:thg-power}
\end{figure*}

\Cref{fig:setup_thg-lambda}(\textbf{d,e}) show the spectral signatures, which were measured using an optical spectrum analyzer (OSA) as the IR detector and a spectrometer as the VIS one. While the IR spectrum (\Cref{fig:setup_thg-lambda}\textbf{a)} displays a very thin line, the VIS one (\Cref{fig:setup_thg-lambda}\textbf{b)}) is broader due to the spectrometer's limited resolution. By scanning the laser's wavelength over a broad range, we can use the traces collected by both the OSA and the visible spectrometer to analyze the overall dependence of the VIS wavelength on the IR one. Note that, while determination of the IR wavelength is straightforward because the OSA peaks are very thin, the limited precision of the spectrometer makes this task a little trickier in the VIS range. Instead of taking the wavelength that maximizes counting, we fit the spectrum to the spectrometer's transfer function,
\begin{equation}
    \sinc^2 [A(\lambda - \lambda_3)] + B
\end{equation}
where $\lambda$ is the VIS wavelength according to the spectrometer, $A$ is a constant related to the spectrometer's resolution, $\lambda_3$ is the THG's wavelength and $B$ represents the constant background noise. \Cref{fig:setup_thg-lambda}~\textbf{c)} shows a very good agreement of the estimated $\lambda_3$ \emph{vs} IR wavelength graph to a linear fit with an angular coefficient equal to $0.33$, consistent with THG.

The power dependence of the collected visible light is characterized using the same experimental setup, only replacing the VIS spectrometer with a power meter. In \Cref{fig:thg-power}\textbf{a)}, we present the collected VIS power as a function of the IR wavelength. As expected from the spectral characterization, several peaks show up (roughly separated by the cavity's free spectral range). This result indicates that the third harmonic peaks are generated from IR modes with neighbouring azimuthal numbers and the same transverse profile, which should be deemed impossible by dispersion. As anticipated by \Cref{eq:energy}, the optical Kerr effect ends up compensating the GVD. Moreover, this compensation leads to appreciable THG for several IR modes of the same transverse family.

The detailed tunability mechanism is the following: as the laser's wavelength is swept, the stored IR energy changes and so do the nonlinear frequency shifts. On one hand, this allows for appreciable stored IR energy over a broader wavelength range. On the other, it allows for the corresponding phase-matched VIS mode to frequency-shift twice as fast as the IR one. If the input power is high enough, at some point these Kerr-driven frequency shifts will be such that doubly resonant THG occurs and appreciable light can be collected at the third-harmonic wavelength.

In order to give a clear characterization of THG, it remains to investigate the question of whether there is a cubic dependence \cite{boydNonlinearOptics2008, butcherElementsNonlinearOptics1990} between input IR and output VIS power. We perform this study for the largest THG peak, near \SI{1546}{\nm}. This measurement cannot be performed at fixed IR wavelength, though, since thermal shifts as well as cross- and self-phase modulation would change the effective coupling between the pump laser and the IR cavity mode, as illustrated by the measurements in \Cref{fig:thg-power}\textbf{(b,c)}. We overcame this issue by sweeping the laser's wavelength and measuring the collected power immediately before the end of the bistability curve of the IR mode, which coincides with the point where the collected VIS power drops to zero, as shown in \Cref{fig:thg-power}\textbf{(d,e)}. This scheme efficiently traces out the effect of nonlinear frequency shifts and results in \Cref{fig:thg-power}\textbf{f)}, where a cubic dependence between collected VIS power and IR pump power is observed.

\begin{figure}[htb!]
	\centerline{\includegraphics[scale=1]{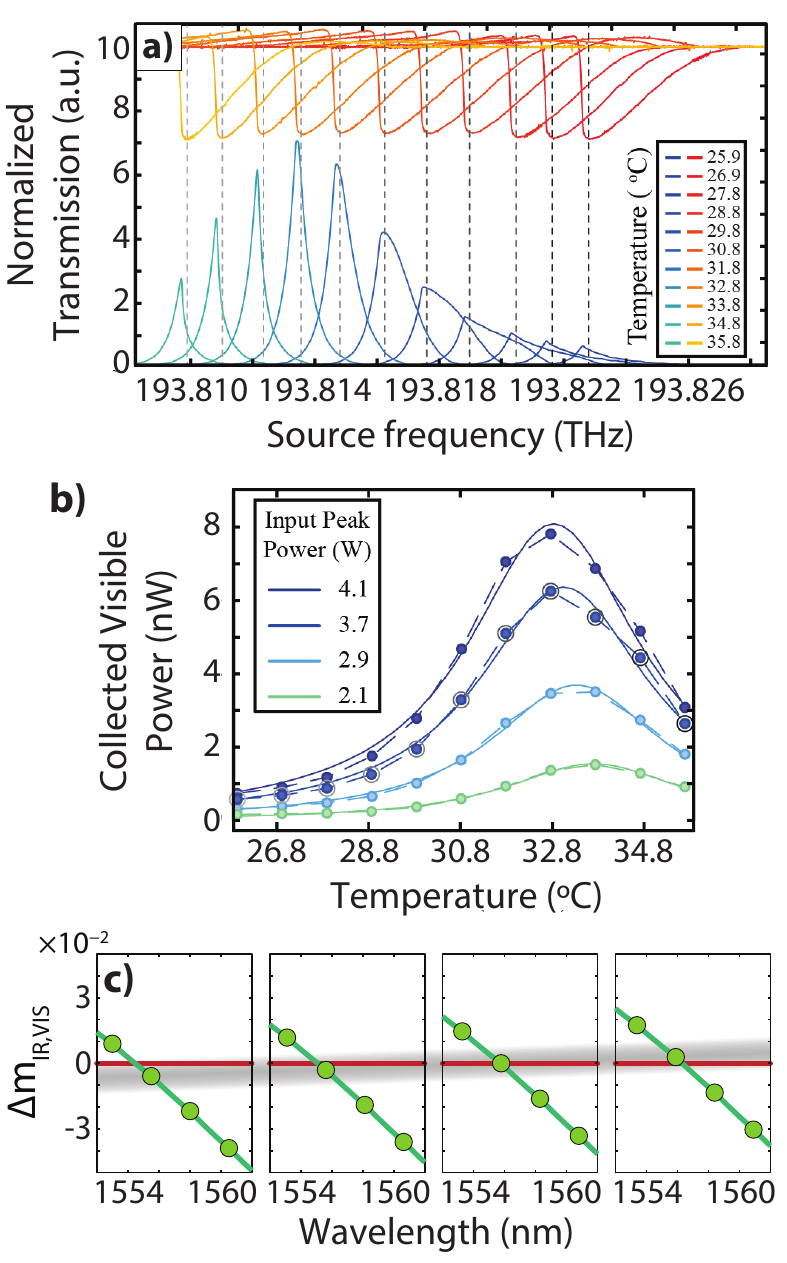}}
	\caption{\textbf{a)} IR transmission (red color scale) and VIS collected power (blue color scale) measured at constant pump power for various temperatures. For each temperature, the thermal detuning is determined as the frequency difference between the lowest IR transmission and the third-harmonic peak.
	\textbf{b)} Visible power as a function of thermal detuning $\delta\omega$ for various values of IR pump power. Dots represent measured data, solid curves are lorentzian fits highlighting the agreement to \Cref{eq:thg-lorentz} when $\delta\omega$ depends linearly on the temperature.
    \textbf{c)} Illustration of the THG maximization scheme. As temperature increases, $\delta\omega$ gets larger and the (interpolated) values of $m_\mathrm{VIS}(\lambda_1/3)$ increasing. For a mode below the phase-matching line, there is an optimum temperature for which phase-matching is perfect.
 }
	\label{fig:thg-temp}
\end{figure}

The collected peak power variation observed in \Cref{fig:thg-power}\textbf{a)} can be better understood from \Cref{eq:energy}. Maximum THG efficiency is achieved if both the IR stored energy $|a_1|^2$ and VIS generated energy $|a_3|^2$ are maximized. This is achieved if
\begin{equation} \label{eq:doubly-resonant}
    \Delta_1 + \alpha |a_1|^2 = 0 = \Delta_3 + 2 \alpha |a_1|^2
\end{equation}
which can be seen as a doubly resonant condition corrected by Kerr-driven frequency shifts. The first part of \Cref{eq:doubly-resonant} fixates the detuning $\Delta_1$ at the power dependent value $\Delta_1^\mathrm{(opt)}$ which maximizes the stored IR energy, $|a_1|^2 = 2 \theta_1 P_\mathrm{in} / \gamma_1^2$. Therefore, the second part of \Cref{eq:doubly-resonant} states that if
\begin{equation} \label{eq:optimum}
    \delta\omega = \omega_3 - 3 \omega_1 = -\frac{4 \alpha \theta_1 P_\mathrm{in}}{\gamma_1^2} - 3 \Delta_1^\mathrm{(opt)},
\end{equation}
then THG is maximized as well. Therefore, the peak collected VIS power drops as $\delta\omega$ deviates from this condition due to GVD. It follows that the largest peak in \Cref{fig:thg-power}\textbf{a)}, near \SI{1546}{\nm}, is the one closest to the optimum condition given by \Cref{eq:optimum} for the input power of that case. Moreover, \Cref{eq:optimum} makes it clear that, if $\delta\omega$ can be controlled, maximum THG can be achieved, in principle, for any pumped IR mode of this transverse family.

Under light of this, we attempt to maximimze THG by placing a pelthier diode under the cavity's chip in order to control its temperature. Because the rate of change of the refractive index with temperature is different at IR and VIS wavelengths~\cite{Toyoda1983}, active control of the cavity's temperature effectively adds a "thermal detuning" $\delta\omega$ between the modes, which influences the phase-matching condition and can greatly enhance or suppress THG according to \Cref{eq:doubly-resonant}. In other words, we are able to compensate for the dispersion using a thermal actuator.

\Cref{fig:thg-temp}~(\textbf{a,b}) show this principle in action, while \Cref{fig:thg-temp}\textbf{c)} gives a pictorial representation of the mechanism. As temperature increases, the thermo-optic effect shifts VIS and IR resonances at different rates, with a net effect of moving the $m_\mathrm{VIS}(\lambda_1/3)$ curve (the same one of \Cref{fig:pm}\textbf{b)}~) upwards. Therefore, a mode slightly under the phase matching condition may get closer and closer to perfect phase matching up to a point where actual phase matching is achieved and THG is optimized. Increasing the temperature beyond this point causes phase matching to be lost and therefore suppresses THG.

In \Cref{fig:thg-temp}\textbf{a)}, we show the measured IR transmission (red tone curves) and the VIS collected power (blue tone curves) for a constant pump power for several temperatures. It becomes very clear how the phase-matching condition is controlled by the temperature, with a very distinct maximum THG efficiency near \SI{32.8}{\celsius}. On the other hand, at $\sim \SI{25}{\celsius}$ the THG can be completely suppressed while the IR resonance is shifted, but otherwise non-affected.

The temperature of maximum efficiency varies with pump power as shown in \Cref{fig:thg-temp}\textbf{b)}, where we plot the collected VIS power for the wavelength fixed near the IR resonance. Since self- and cross-phase modulation also displace the IR and VIS resonances, they influence the phase-matching condition differently at different input power levels. Nevertheless, temperature control may still compensate for these nonlinear shifts as evidenced by the appearance of the well defined peaks at different thermal detunings in \Cref{fig:thg-temp}\textbf{b)}. This result also agrees with \Cref{eq:optimum}. It is worth noting how these curves agree well to a Lorentzian fit, exactly as expected from \Cref{eq:energy}. In fact, using the definition of $\delta\omega$ from \Cref{eq:optimum} in the second of \Cref{eq:energy},
\begin{equation} \label{eq:thg-lorentz}
    |a_3|^2 = \frac{g^2 |a_1|^6}{(3 \Delta_1 + 2\alpha |a_1|^2 + \delta\omega)^2 + \gamma_3^2}
\end{equation}
which explains the observed Lorentzian lineshape as long as $\delta\omega$ is proportional to the temperature as long as the IR detuning and input power (and hence the stored IR energy, $|a_1|^2|$) are kept constant.

\section{Conclusion}

We demonstrate tunable third-harmonic generation in high-Q wedge silicon dioxide microcavities. Tunability is achieved due to the tuning of the phase-matching condition for several pairs of infrared and visible modes, which is provided by the cavities own nonlinear frequency shift mechanisms. While the typical wavelength relation between input and third-harmonic wave was readily verified, a clever approach to separate THG from nonlinear frequency shifts was devised and successfully applied.

Moreover, we show that any residual deviation from the phase-matching condition can be compensated by changing the pumped and third-harmonic resonance frequencies, which was performed by controlling the cavity's temperature. Finally, we show how this third-harmonic enhancement strategy agrees reasonably well with coupled-mode-theory. Therefore, our demonstration establishes an  exciting route for integrated silicon-based devices for generation of tunable coherent light sources in the visible range through third-harmonic generation.


\section*{Funding}
This work was supported by S\~{a}o Paulo Research Foundation (FAPESP) through grants 
2015/22178-1, 
2016/05038-4, 
2018/21311-8, 
2013/06360-9, 
2012/17765-7, 
2018/15580-6, 
2012/17610-3, 
2018/15577-5, 
2018/25339-4, 
Conselho Nacional de Desenvolvimento Cient{\'i}fico e Tecnol{\'o}gico (CNPq),
Coordena{\c c}\~ao de Aperfei{\c c}oamento de Pessoal de N{\'i}vel Superior - Brasil (CAPES) (Finance Code 001), and Financiadora de Estudos e Projetos (Finep).

\section*{Acknowledgments}
The authors thank A. Von Zuben for technical support. We also acknowledge CCSNano-UNICAMP for providing part of the microfabrication infrastructure.

\section*{Disclosures} 
The authors declare no conflicts of interest.

\section*{Data availability}
Data underlying the results presented in this paper are available upon reasonable request. 


\end{document}